# PROOF-OF-VALUE-ALIENATION (PoVA).

Tim Shuliar [6u66le9um@gmail.com], Nikita Goldsmit [sjalfa777@gmail.com].

## Table of Contents

# PREFACE.

In this paper, we will describe a concept of a cryptocurrency issuance protocol which supports digital currencies in a Proof-of-Work ("PoW") like manner. However, the methods assume alternative utilization of assets used for cryptocurrency creation (rather than purchasing electricity necessary for "mining").

The scope of this paper is relevant to the economic properties of crypto assets and their valuation, rather than tech or cryptography issues.

The proposed concept may be executed in various forms and on different platforms - based on blockchain or even centralized networks. We are focused on the economic aspect of value creation and source of the scarcity of new asset which *may* be realised on blockchain with no human control and *may* equally have a centralised verification of its issuance, as shown below.

# PROOF-OF-WORK.

Value of fiat currency is a function of its adoption (representing demand) and scarcity (representing supply).

Fiat money is a currency without intrinsic value that has been established as money, often (*but not necessarily*) by government regulation. Fiat money does not have use value and has value only because a government maintains its value, *or because parties engaging in exchange agree on its value*. Most general purpose (not utility or debt tokens) cryptocurrencies fall under the definition of fiat currency.

Value of cryptocurrencies is determined by supply and demand, with supply being artificially limited by design of the source code of most cryptocurrencies.

Early PoW based cryptocurrencies, such as Bitcoin and Ethereum retain the highest capitalization among all crypto assets (as of the time of writing of current paper). Apparent reasons for that are predictable declining supply of new coins, relatively predictable increasing production cost of new coins and steady demand (determined by belief that these assets will appreciate in the future and value of cryptocurrencies as a means of exchange for certain transactions).

We mention relative predictability of the self-price because one can generally be sure it is increasing over time, but the increase is not linear or say uniformly accelerated, but rather raising on average with variable (positive and negative acceleration) rate. The reason behind the uncertainty and non-uniformity is a large number of factors affecting unit mining self-price, including but not limited to the price of electricity, cost of mining equipment, development of specific mining equipment, such as ASICs, etc.

Networks supporting PoW cryptocurrencies need to have adjustable mining difficulty in order to support predictable supply rate. The difficulty is adjusted periodically as a function of how much hashing power has been deployed by the network of miners, allowing a cryptocurrency to have a predictable declining production rate. In case of Bitcoin every 2,016 blocks (approximately 14 days at roughly 10 min per block), the difficulty target is adjusted based on the network's recent performance, with the aim of keeping the average time between new blocks at ten minutes.

We argue that all PoW backed cryptocurrencies are backed by "real" fiat money spent on mining in a form which is very similar to Proof-of-Burn (as described below). In case of mining, fiat money is spent on necessary electricity and equipment. On the technical and cryptographic level, PoW cryptocurrencies are issued against a provable mathematics «work». From the financial standpoint of view same tokens are issued against the (fiat) investment in equipment and electricity, of which prices are financially restrictive factors for both, rapid depreciation of such currencies and several forms of attacks (such as Sybil attack). This «burning» (of fiat money) is indeed the factor which assigns a value to the PoW tokens.

For the sake of argument let's imagine that one has access to «free» equipment and electric power. Such actor would be able to compromise and/or control the network. In case if all users would gain similar access to "free resources", price of mined cryptocurrency would rapidly drop to zero, as it would be possible to produce it at no cost.

Since no new units of the currency may be produced cheaper than the minimum ever-increasing threshold and none of the miners would be incentivised to sell mined coins at a price lower than mining cost - an artificial minimum price of the coins is being established.

So, in our view source of the scarcity of PoW cryptocurrencies come from the inevitable necessity of fiat burning to produce new units. Even though we can't say that PoW cryptocurrencies are backed with fiat money directly since there is no guarantee of reverse conversion, we are safe to say that their scarcity is guaranteed with proof of burning of fiat money, and minimum value of issued tokens is secured by the same process.

In our view it is possible to create a general-purpose cryptocurrency, using existing (or new) blockchain, by issuing it against alienation of value (in fiat or crypto form) and direct that value for useful causes. We shall describe the concept below.

## PROPOSED CONCEPT.

We propose a new protocol of issuance of cryptocurrencies based on proof of value alienation in the form of non-compensated (other than with new crypto tokens) spending.

Initially, we planned to use a term Proof-of-Spending. However, we found out that it has been already used in several cases [one[1], two[2]] and realised that spending itself assumes receiving compensation for spent funds, which is unacceptable for the proposed concept. Some other, more complicated concepts, which assume minting of tokens at the moment charitable spending is performed, such as Charityblockchain[3] are also significantly different from the proposed concept. In mentioned example donors receive bonuses tied to fiat currencies, making it more of a cashback-like system, while speculate-able cryptocurrency units issued against the donation are not directed to donors and used for other purposes.

If money is spent to purchase goods or services or obtain anything of equal value, and new units of currency is issued against such action than we consider it a form of cash back (if the purchase was discounted compared to normal price) or non-backed issuance (if the cryptocurrency issuance does not affect the deal price). Both cases are out of the scope of the proposed concept and have no relevance to our proposition except for a naming of protocols, which may be confusing. We shall consider issuance against fully compensated spending a non-backed issuance as such issuance has no relevance to the act of spending, and while two parties (buyer and seller of goods) exchanged value, no new value has been created. Thus, token issued in such fashion (in our view) has no value whatsoever.

---

[1] https://arxiv.org/abs/1804.11136v4

[2] https://ccrb.io/

[3] https://charityblockchain.io

The transaction could've taken place with or without such issuance, and such issuance could've occurred just as well without transaction itself. Fact of transaction itself in our view may not be a basis for the creation of value, and tokens derived from such spending may not be viewed as money. With the same success, one may start issuing coins whenever wind speed reaches certain threshold, or GDP of USA reaches certain value, or any other unrelated process, which does not assign a value to the newly issued token.

Non-compensation spending, or value alienation, on the other hand, does indeed assign value to a token issued against it, as it limits issuance rate and secures a minimum price of the token, because holder of the token will be reluctant to liquidate it at a price lower than what was spent on its production (we are not taking into consideration extreme situations and exclusions from the generally reasonable behaviour).

Financial consequences for the issued currency under our proposition are similar to those of Proof-of-Burn[4] concept proposed by Iain Stewart[5] (with a notable difference in the financial consequences for the underlying asset, which is being alienated from the holder, as described below). Proof-of-Burn is a method for distributed consensus and an alternative to Proof-of-Work and Proof-of-Stake. It can also be used for bootstrapping one cryptocurrency off of another.

The idea is that miners should show proof that they burned some coins - that is, sent them to a verifiably non-spendable address. This is expensive from their point of view, just like Proof-of-Work; but it consumes no resources other than the burned underlying asset.

A first notable difference between proposed concept and Proof-of-Burn is that under proposed concept funds used to issue new tokens, shall not be «burned» or blocked forever, but rather be utilized for public good. This may be achieved in various forms, the only necessary condition is that «miners» shall not directly benefit from alienated value, as it would create a "loop of value", preventing the creation of value.

The second notable difference is that our proposal may be realized with fiat money, rather than cryptocurrency, which might be highly beneficial for certain causes, as shown below.

One of the possible implementations is to virtualize mining (which may be realized in the form of virtual «hashrate» (1) or any other simulated measurable proxy index, produced by virtual «mining rigs».

---

[4] https://en.bitcoin.it/wiki/Proof_of_burn

[5] https://en.bitcoin.it/wiki/User:Ids

Other implementation - direct issuance (2) against proved alienation of value). Various proofs of value alienation may be realized; we list some of the most obvious below under the «types of proofs» section of the current paper. Funds used to issue new currency may be aggregated and directed to public good, as described below.

In first (1) case holders of virtual «mining rigs» will be obtaining new units of issued cryptocurrency throughout the time they hold said virtual «rigs». Virtual rigs may have a flat or declining efficiency. Such rigs may be realised in the form of a smart contract, guaranteeing further issuance. However, the issuance rate will be dependent on the amount of total «miners». This might seem to be an over-complication of the process, but for certain use cases, such a setup might be preferable. Some of the most obvious use cases of the proposed protocol of issuance described below, under «use cases» section of the current paper.

Besides continuation of the use of «virtual rigs» might require additional payments to a particular wallet, for a smart contract to continue issuing new tokens. Such a scenario may be viewed as an analogue to paying for electricity bills while mining PoW cryptocurrencies.

The second case (2) is more straightforward and may be realised as a direct issuance against non-compensated (otherwise than with issued tokens) spending. Collected funds should not directly or immediately benefit spenders, but rather be used for public benefit or benefit of individuals or groups nonaffiliated to spenders.

# TYPES OF PROOFS.

A simplest and most apparent PoVA based system may be realized using a smart contract on one of the Turing complete blockchains. As this solution is relatively straightforward, we shall not give too much attention to its description. For the sake of simplicity, we shall call an individual who performs an act of alienation of value a virtual miner.

A sequence would be as follows:

1. A virtual miner transfers any amount of accepted coins to issuer-DAO. For the sake of simplicity, we shall call such coins Prime Assets.
2. Issuer-DAO under a smart contract issues a relevant number of tokens (we shall call those Derivative Tokens) and transfers them to virtual miner's wallet.

3. Under the smart contract terms, DAO transfers received Prime Assets to charity, an investment vehicle, into the countries or organisation budget, or to other beneficiaries (as described under the use cases section), depending on the scenario of PoVA protocol usage.
4. After a particular time, (or depending on the change of another factor, such as the total amount of issued coins) issuance ratio is adjusted (see the source of scarcity section).

Another possible option is issuance based on virtual mining. In this case one of the possible sequences would be as follows:

1. A virtual miner transfers any amount of Prime Assets to issuer-DAO.
2. Issuer-DAO includes virtual miner's wallet into a smart contract.
3. Under this smart contract, issuer-DAO shall periodically issue and transfer a certain number of Derivative tokens to the virtual miner's wallet. A particular number of tokens may be a function of variables, for example, total virtual miners participating in the process or the total amount of Prime Assets invested.
4. Under the smart contract terms, DAO transfers received Prime Assets to charity, an investment vehicle, into the countries or organization budget, or to other beneficiaries (as described under the use cases section), depending on the scenario of POVA protocol usage.

Such principle is closer to classic PoW cryptocurrencies, as certain number of units will be issued per unit of time under any circumstances.

The same principle may be implemented in various forms, e.g. virtual miner holds the cryptographic key, verified by a smart contract to receive Derivative coins. Another variation assumes necessity to make regular transfers of Prime Assets (analogue to paying for electricity under PoW). There are maybe other realizations, and they are out of the scope of current paper - our goal is limited to the high-level description of the core concept.

Most important: the core economic principle remains the same in case if we use fiat money as Prime Asset. In this case one will be unable to use an entirely smart contract-based process, however, for some applications, there are suitable ways to verify fiat money transfer.

For example, arrangements may be achieved with either SWIFT, ACH (for US money transfers) or even particular bank, to provide an API for automatic information exchange regarding specific transfers. Otherwise, certain application with the same value creation principle may include an element of trust to authority, or organisation, such as charitable, religious or governmental organisation. Such a concept is entirely different from «classical» crypto community values but might still have a value for certain other ideological groups. Economic meaning of value assigned to a new token remains the same.

## SOURCE OF SCARCITY.

Supply of coins issued under the PoVA protocol may be artificially scarce and ever-decreasing, promoting an ever-growing production cost of the coins. Shall the demand for the coin be raising or constant, the price of the coin will be continuously increasing.

Such a decrease of supply (or an increase of cost price) should be predictable and known to the public. In most PoW cryptocurrencies supply rate of coins is constant, but production cost is ever increasing.

In PoVA case increase of production cost may be directly defined by a mathematical function, e.g. logarithmic regression. Otherwise, the rate of decrease of supply may be a function of total participants of the system (virtual miners), the total influx of Prime Assets, or other factors.

## USE CASES.

PoVA obviously may have many different applications, and it's out of the scope of current paper to find as many of them as possible. We will mention several simplest and most obvious applications where described concept might add value.

1. Community-based cryptocurrency. Any large community might have an incentive to produce its means of payment and value storage. It might be a religious group, group sharing the certain political belief, or even a nation. PoVA might be used in a various form - starting from fully autonomous blockchain / smart contract-based solution, all the way up to the situation when a central authority (central bank) issues against direct payment or sells pre-issued coins to the public.
2. Various forms of Charitable Foundations. In this case, Alienation of Value of Prime Asset shall be done in the form of donation. Thus donor might have two different motives - to perform a donation or to «mine» a Derivative coin, or combination of both.

3. Ideological cryptocurrency. In such situation, all the Prime Assets may go into a fund, which will either invest on behalf of backers, or simply hold assets, and use for further (distant in time) benefit of the «miners». Such fund may provide pension-like bonuses, or UBI (universal basic income) solution - any non-direct or distant in time benefit would be potentially applicable.
4. General purpose cryptocurrency, which is not focused on the use of Prime Assets. For example, one can imagine a smart contract which would pay collected Prime Assets (in the form of cryptocurrency) out to a random «miner» periodically. This would be a lottery type of situation. However, the value assignment process to a Derivative token would remain the same.

We believe that POVA will find its place in a world of alternative assets and will create value for specific commercial or nonprofit projects in the future. Our goal was to describe the economic aspect of value creation in the form of new crypto asset and propose a concept of how this process might be changed, to use Prime Assets for public benefit.

# REFERENCES.